\documentclass[fleqn,usenatbib]{mnras}

\usepackage{newtxtext,newtxmath}

\usepackage[T1]{fontenc}

\DeclareRobustCommand{\VAN}[3]{#2}
\let\VANthebibliography\thebibliography
\def\thebibliography{\DeclareRobustCommand{\VAN}[3]{##3}\VANthebibliography}

\usepackage{graphicx}	
\usepackage{amsmath}	

\usepackage{ulem}

\newcommand{\dd}{\mathrm d}

\title[Massive star cluster origin for the GCR population]{Massive star cluster origin for the galactic cosmic ray population at very-high energies}

\author[T. Vieu and B. Reville]{
T. Vieu,$^{1}$\thanks{E-mail: thibault@mpi-hd.mpg.de}
B. Reville,$^{1}$
\\
$^{1}$Max-Planck-Institut f\"ur Kernphysik, Saupfercheckweg 1, D-69117 Heidelberg, Germany
}

\date{Accepted XXX. Received YYY; in original form ZZZ}

\pubyear{2022}

\begin{document}
\label{firstpage}
\pagerange{\pageref{firstpage}--\pageref{lastpage}}
\maketitle

\begin{abstract}
We demonstrate that supernova remnant (SNR) shocks embedded within massive star clusters can reproduce both the cosmic-ray proton and all-particle spectra measured in the vicinity of the Earth up to hundreds of peta-electronvolts (PeV). We model two classes of massive star clusters. The first population are ``loose clusters'' which do not power a collective wind termination shock. SNR shocks then expand in a low-density and weakly magnetised medium, and this population mainly contributes up to the ``knee'' of the CR spectrum around 1~PeV. The second population are young compact clusters, which are powerful and compact enough to sustain a collective wind outflow. SNR shocks then expand from the cluster into the strongly magnetised wind and accelerate nuclei up to ultra-high energies. This population, representing only about 15\% of all Galactic massive star clusters, nevertheless dominates the spectrum between $\sim$ 1 and 100~PeV. While these two components alone can reproduce the shape of the CR spectrum up to hundreds of PeV, adding a light sub-ankle extragalactic component motivated by composition and anisotropy measurements, allows to reproduce the spectrum up to the highest energies. Fitting parameters are systematically linked to physical variables whose values are in line with theoretical expectations.
\end{abstract}

\begin{keywords}
cosmic rays -- acceleration of particles -- open clusters and associations: general -- shock waves
\end{keywords}



\section{Introduction}
Several arguments suggest that the transition between galactic and extragalactic cosmic rays (CR) begins around several hundreds of PeV, with a subdominant extragalactic component below 1~EeV, which becomes dominant at the ``ankle'' at about 3~EeV. For example, measurements of the CR composition by several experiments, despite uncertainties in the modelling of hadronic air showers, show that the mean logarithmic mass of the detected particles increases with energy beyond about 1~PeV, peaks at several hundreds of PeV, and then decreases up to the ankle, where it starts to increase again \citep[see Figure~9 in][]{supanitsky2022}. This is consistent with a scenario in which CRs are accelerated within galactic sources with successive charge dependent cut-off, resulting in an excess of heavy elements around the maximum energy of these galactic sources, before a light sub-ankle component starts to take over.
This picture is supported by measurements on the dipole anisotropy, which show that the phase of the anisotropy starts to smoothly increase towards the Galactic Anti-centre beyond 100~PeV \citep[see Figure~9 in][]{Kachelriess2019}. The modest amplitude of the anisotropy in these energy bands disfavours a galactic origin for the light CR component in this energy range \citep{giacinti2012}. Galactic sources thus struggle to explain simultaneously composition and anisotropy measurements beyond 100~PeV, while the measurements are consistent with an extragalactic scenario. Moreover, at hundreds of PeV the Larmor radius of CRs in galactic magnetic fields becomes comparable to the scale-height of the Galactic Disk, which further disfavours a galactic origin at higher energies due to confinement issues.  Well below the ankle, it has been pointed out \citep[e.g.][]{parizot2014} that a concave feature separating two extended power-laws, such as the ``knee'' observed in the CR spectrum at about 3~PeV, requires fine-tuning to be accounted for in a two-component model, with the second component being subdominant yet non-negligible. If the two components are completely unrelated, as is the case if one assumes that the CRs below/above the knee are of galactic/extragalactic origin, such tuning is unlikely and a smooth spectrum is not expected in principle. On the other hand, the second knee  at about 0.5~EeV, which is a less pronounced concave feature, suffers less from this issue as the steep power-law component extends beyond it over at most a decade, and therefore this feature could mark the turn-over of the heavy galactic CRs, a few hundreds of PeV before the light extragalactic component takes over \citep[e.g.][]{thoudam2016}, especially if the latter is not a perfect power-law, as suggested by a number of models \citep[e.g.][]{unger2015,globus2015,Kachelriess2017,Berezinsky2006}.

Setting the transition between galactic and extragalactic CRs at hundreds of PeV implies that at least a subset of the galactic sources -- on the order of 20\% \citep{Axford1994,parizot2014} -- must be able to accelerate heavy nuclei up to hundreds of PeV, i.e. protons beyond 10~PeV. Identifying these sources is not a straightforward task and there is no consensus on the matter at the present time.
Indeed, it has long been suspected that supernova remnant (SNR) shocks expanding in the warm interstellar medium (ISM) struggle to reach PeV energies \citep{lagage1983}, even when considering CR-induced instabilities in the dense wind blown by the progenitor star \citep{bell2013}, unless the SNR shock is exceptionally fast or exhibit extreme transport properties \citep[e.g.][]{voelk1988,bell2001,ptsukin2010,MakotoKirk15}. However, even in the most optimistic scenarios, the maximum energy is an order of magnitude lower than what is required to account for the GCR spectrum up to hundreds of PeV.

A number of alternative hypotheses accounting for the GCR spectrum in the range $1-100$~PeV have been proposed, including, e.g., the re-acceleration of particles by the galactic wind termination shock \citep{Jokipii1987,Zirakashvili2006,merten2018}, the re-acceleration by multiple scatterings in the Fermi bubbles \citep{cheng2012}, the collective effects in superbubbles \citep{bykov2001b,bykov2018}. It is as well possible that a different class of sources takes over beyond the knee, e.g. the Galactic Centre \citep{HESS2016}, microquasars \citep{BoschRamon2005}, hypernovae \citep{Sveshnikova2003}, or young massive stars \citep{biermann1993} and young stellar clusters blowing a collective wind \citep{aharonian2019,vieu2022Emax}. The interest towards the latter has been rapidly raising in the recent years since the discovery that these sources are efficient gamma-ray emitters up to hundreds of TeV \citep[e.g.][]{gammaCygnus_HAWC2021,cao2021,HESSWd12022}.

It was realised early on \citep[e.g.][]{higdon2005} that most massive stars (and hence SN progenitors) form in clusters. The powerful stellar feedback of the young early-type stars subsequently carve hot ``superbubbles'' within the parent molecular cloud \citep{weaver1977,mccray1987}. Efficient particle acceleration is expected to take place within these  environments, by the stellar winds \citep{montmerle1979,cesarsky1983}, by the turbulence \citep{bykov2001,klepach2000}, or by supernova explosions \citep{parizot2004,ferrand2010,vieu2022}. Several arguments support a superbubble origin of CRs; in particular, the enrichment of the accelerated medium by the winds of Wolf-Rayet stars has been shown to resolve the $^{22}\rm{Ne}/^{20}\rm{Ne}$ anomaly of the GCRs \citep{gupta2020}.

In a recent paper \citep{vieu2022Emax}, we assessed the maximum proton energies achievable under various particle acceleration scenarios around massive star clusters. In particular, we have shown that, similarly to the case of isolated SNR shocks, it is not straightforward to reach PeV energies when realistic environmental parameters are considered. Nevertheless, we highlighted a promising possibility, that is to accelerate particles around a SNR shock launched from a compact cluster into the collective large-scale ($\sim 10$~pc) wind blown by the massive stars. Large magnetic fields are amplified in the core of the cluster, potentially up to hundreds of \textmu G, and although the level of magnetic turbulence decreases in the powerful collective wind, the average magnetic field in the region swept up by the SNR shock is still expected to be about one order of magnitude larger than the average interstellar value, which can facilitate effective particle acceleration. Under reasonable assumptions on the efficiency of turbulence generation inside the cluster, we computed the maximum energy of the particles as $E_{max} \lesssim 5 Z u_5$~PeV for standard cluster parameters, where $u_5$ is the initial SN shock velocity in units of 5000~km/s. This suggests that, in favourable conditions, a fast ($>10$~000~km/s) SNR shock can accelerate protons beyond 10~PeV, and the fastest ($>30$~000~km/s) shocks could almost accelerate iron up to 1 EeV, therefore meeting the required bounds to explain the GCR spectrum up to hundreds of PeV.

In the present work, we aim to compute the GCR spectrum including this component due to particle acceleration around this special population of SNR shocks. We start in Section~\ref{sec:MSCpopulation} by a census on the galactic star-cluster population, which is used to estimate the frequency of these SN explosions. Section~\ref{sec:spectraSN} is devoted to a description of a two-component model, including the contribution of SNR shocks expanding in loose associations of massive stars, where no collective wind termination can form, as well as the contribution of clustered SNR shocks. We eventually compute the overall contribution of massive star clusters to the GCRs in Section~\ref{sec:galspectrum}, where we discuss the agreement between our model and measurements made in the vicinity of the Earth between 1~TeV and 1~EeV. We conclude in Section~\ref{sec:conclusions}.

\section{The massive star cluster population of the Galaxy}\label{sec:MSCpopulation}
In this Section we review and categorise the galactic cluster census presented in the recent analysis of the Gaia Data Release 2 (DR2) \citep{Cantat2020}.

\subsection{Definitions}\label{sec:MSCpopulationdef}
Because star clusters are at the crossroads of several sub-fields of astrophysics, e.g. cosmic-ray physics and astrometry, the nomenclature is not well-defined. In the present work, we use the following designations, which categorise massive star clusters in a way which will prove useful in what follows.

\textbf{\textit{Star cluster}}: a star cluster is defined as an association of stars over a scale under a few tens of pc, which is not necessarily gravitationally bounded. These clusters are referred to as \textit{open clusters} in the astrometry community, as opposed to \textit{globular clusters} which are much older and extend to much larger scales. We do not presently discuss globular clusters as they are not expected to represent a sizeable fraction of the GCR sources.

\textbf{\textit{Massive star cluster (MSC)}}: a massive star cluster is a star cluster which contains SN progenitors, i.e. stars with zero-age main-sequence (ZAMS) mass larger than about 8$M_\odot$ \citep[e.g.][]{Poelarends2008,Smartt2009}\footnote{In the following we disregard the fact that some stars with ZAMS mass larger than $8M_\odot$ might not explode \citep[e.g.][]{sukhbold2016}.}. Importantly, this excludes all clusters which are more than about 40~Myr old, as all SN progenitors are expected to explode within 40~Myr, assuming coeval star formation \citep{limongi2006}.

\textbf{\textit{Young massive star cluster (YMSC)}}: a young MSC is a MSC which still contains stars expected to undergo a transition to the WR phase in the end of their life. Assuming roughly that stars with ZAMS mass larger than 20$M_\odot$ are expected to become WR stars \citep{Crowther2007,Sander2012}, young clusters are clusters which are less than about 10~Myr old, assuming coeval star formation.

\textbf{\textit{Evolved massive star cluster}}: in contrast to YMSCs, evolved MSCs are defined as MSCs which do not contain WR stars anymore (i.e. which are more than $10$~Myr old). Their wind mechanical power is then much smaller than that of YMSCs, such that their energy budget is largely dominated by SN explosions \citep{vieu2022}.

\textbf{\textit{Compact massive star cluster}}: in the present work, we consider that a MSC is compact if its half-mass radius is lower than $3 (t_{\rm c}/\mathrm{Myr})^{2/5}$~pc, where $t_{\rm c}$ is the cluster age. The physical motivation lying behind this somewhat arbitrary threshold is clarified below.

\textbf{\textit{Extended massive star cluster}}: in contrast to compact MSCs, extended MSCs are MSCs whose stellar distribution extends beyond $3 (t_{\rm c}/\mathrm{Myr})^{2/5}$~pc.

\textbf{\textit{Wind-blowing cluster (WBC)}}: if a YMSC is compact enough, the powerful winds of the young embedded early-type and Wolf-Rayet (WR) stars are expected to merge into a collective outflow. This collective wind terminates at the wind termination shock (WTS), whose radius can be estimated theoretically as \citep{weaver1977}:
\begin{equation}\label{CR_in_SB:radiusTS}
R_{\rm WTS} \approx N_*^{3/10} \left( \frac{n_{\rm ext}}{100~{\rm cm}^{-3}} \right)^{-3/10}  \left(\frac{t_{\rm c}}{1~{\rm Myr}} \right)^{2/5} ~{\rm pc}
\, ,
\end{equation}
where $N_* \approx 10 - 1000$ is the total number of massive ($M\gtrsim 8 M_\odot$) stars in the cluster, $n_{\rm ext} \approx 1 - 100$~cm$^{-3}$ is the density of the medium beyond the superbubble shell, $t_{\rm c} = 1 - 10$~Myr is the age of the (young) cluster. In order to express the WTS radius as a function of the number of massive stars $N_*$, we assumed an average mechanical power per star of $10^{36}$~erg/s, which, as will be discussed in details in Section~\ref{sec:stellarwinds}, is only expected in the case of a young cluster. If the clustered massive stars are distributed over a region smaller than $R_{\rm WTS}$, then the collective wind can form. Otherwise, the theoretical model breaks down and the individual winds are not expected to merge in a collective outflow. We conveniently define wind-blowing clusters as the subset of YMSCs which are compact enough to power a collective wind beyond the boundary of the cluster. Eq.~\ref{CR_in_SB:radiusTS} provides an estimate of $R_{\rm WTS}$ ranging from about $2 (t_{\rm c}/\mathrm{Myr})^{2/5}$~pc for small clusters to about $20(t_{\rm c}/\mathrm{Myr})^{2/5}$~pc for the most powerful star clusters evolving in a rather dilute environment. The strong dependency of the WTS size on the cluster and environmental parameters, as well as the difficulty to link the theoretical ``cluster boundary'' to the observable ``half-mass radius'', or the difficulty to estimate the total number of massive members from observations, make it challenging to set a simple compactness threshold for the formation of a WTS. In the following, in order to constrain the fraction of WBCs, we will assume a theoretical WTS shock radius of $3 (t_{\rm c}/\mathrm{Myr})^{2/5}$~pc, as the average cluster contains a few tens of massive stars and evolves in a dense external medium. We consider this to be a conservative value and can be understood as an upper bound, which only has a meaning when averaged over a large number of MSCs.

\textbf{\textit{Loose cluster (LC)}}: in contrast to the population of WBCs, we define loose massive star clusters as the MSCs which are unable to power a collective outflow, which includes both extended and evolved clusters.

\subsection{Massive cluster population as inferred from Gaia DR2}
\begin{figure*}
          \centering
              \includegraphics[width=\linewidth]{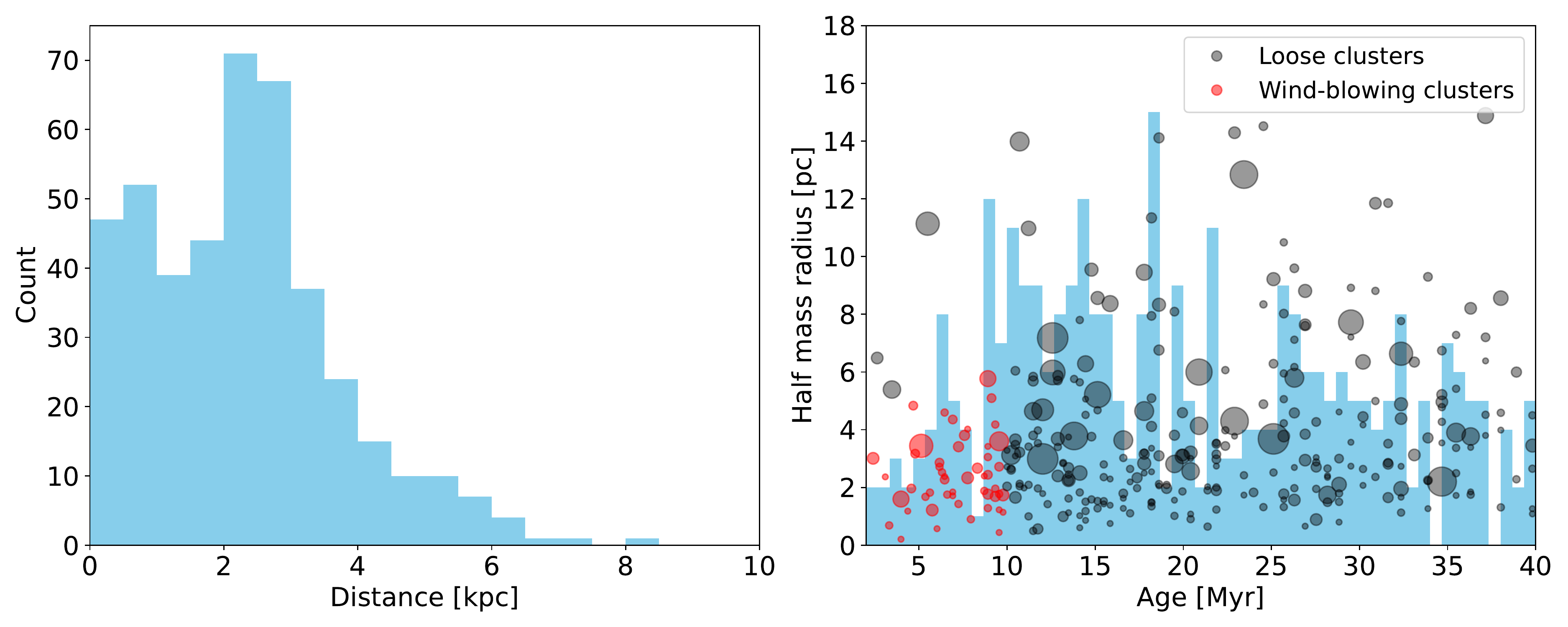}
   \caption{Left: Radial distribution (centred at the Sun) of the MSCs extracted from the open cluster catalogue of \citet{Cantat2020}. Right: MSCs located within 3~kpc, classified in two categories: wind-blowing clusters (red disks) and loose clusters (grey disks). The size of the disk is proportional to the number of members detected in the cluster. The histogram in the background shows the age distribution of the clusters.}
   \label{fig:Gaia_DR2_OC}
 \end{figure*}

\citet{Cantat2020} have performed a systematic analysis of the clusters identified in the Gaia DR2. They used artificial neural network techniques in order to characterise 1867 star clusters distributed within a disk of radius about 6~kpc around the Sun. They estimated, among other properties, the age of the clusters, their half-mass radii and their stellar content with membership probability larger than 70\%. As particle acceleration is expected to be predominantly powered by massive stars, we restrict the sample to the massive star clusters, which are less than 40~Myr old. This represents a total of 430 MSCs. As shown in the left panel of Figure~\ref{fig:Gaia_DR2_OC}, the distribution of these clusters as a function of distance from the Sun drops beyond 3~kpc, due to the difficulty of inferring reliable cluster parameters from reddened and blurred colour-magnitude diagrams observed in regions of high extinction. In order to extract a representative sample of the cluster distribution in the Galaxy, we disregard in the following the tail of the distribution beyond 3~kpc. We are then left with 320 ``nearby'' MSCs located within and in-between the Local, Perseus and Sagittarius arms.

Knowing the half-mass radii and the ages, we can classify the clusters in our sample according to the definitions previously set. The result is shown on the right panel of Figure~\ref{fig:Gaia_DR2_OC}. Although most of the clusters are compact, with a half-mass radius $\lesssim 5$~pc as seen in Figure~\ref{fig:Gaia_DR2_OC}, only 50 are young enough to be wind-blowing clusters. These numbers are summarised in Table~\ref{tab:clusterpopulation}.

The histogram showing the distribution of MSCs as a function of their age, which is plotted in the background of the right panel of Figure~\ref{fig:Gaia_DR2_OC}, does not display any correlation. This is expected since the lifetime of the massive members in a star cluster is much smaller than the cluster disruption time \citep{krumholz2019}.

\begin{table}
\centering
\begin{tabular}{lcc}
\hline
& Young & Evolved  \\
Compact & 16\% & 80\% \\
Extended & 1\%  & 3\%  \\
\hline
\end{tabular}
\caption{Representation of massive star clusters in the catalogue of \citet{Cantat2020} located within 3~kpc from the Sun, classified according to the definitions established in Section~\ref{sec:MSCpopulationdef}.}
\label{tab:clusterpopulation} 
\end{table}

\subsection{On the difficulty of estimating the number of massive star clusters in our Galaxy}
One can try to estimate the number of MSCs expected in the Galaxy on theoretical grounds, considering an average star formation rate of $2 M_\odot$/yr \citep{molinari2014} and a clustering fraction of 80\% \citep{higdon2005}. This results in a total stellar mass of about $M_{\rm tot} = 3 \times 10^7 M_\odot$ over 20~Myr -- the average MSC age.
The sample of observed MSCs are distributed according to an IMF scaling as $M_{\rm cl}^{-2}$ \citep{krumholz2019}, such that the total number of MSCs in the Galaxy reads:
\begin{equation}\label{numberclusters}
    N_{\rm cl} = \frac{M_{\rm tot}/M_{\rm min}}{\ln \left(M_{\rm max}/M_{\rm min} \right)} \, ,
\end{equation}
where $M_{\rm min}$ ($M_{\rm max}$) are the lower (higher) ends of the MSC distribution in the Milky Way.
The maximum mass of a star cluster (typically $10^5 M_\odot$) only enters the logarithm in Eq.~\ref{numberclusters} and is therefore not a critical parameter, in contrast to the minimal mass, which must be precisely specified, although it is not well-defined physically. Indeed, the $M_{\rm cl}^{-2}$ distribution extends down to very low masses, e.g. $10^2 M_\odot$ \citep[][]{lamb2010}, but it is not relevant to talk about a ``massive'' star cluster if it only hosts one or two massive members. If $M_{\rm min}$ varies in the range $10^2 M_\odot - 10^3 M_\odot$, one obtains $N_{\rm cl} \approx 6000 - 45000 $, i.e., assuming that these are distributed over a galactic disk of radius 15~kpc and that the local star formation rate is half that of the Galaxy average \citep{miller1979}, we should expect 100 to 1000  MSCs in the 3~kpc vicinity of the Sun. The number of 320 clusters catalogued in \citet{Cantat2020} lies within this range and we could extrapolate it to a total of about 16k clusters in the Galaxy, which would be obtained from Eq.~\ref{numberclusters} assuming a minimum mass of $450 M_\odot$ (6 massive members in the cluster). This appears to be a reasonable estimate {\it a posteriori}, though the completeness of the local cluster catalogue is far from certain. Finally, given that the star formation rate is inhomogeneous on the scale of the entire Galaxy, one might wonder if these extrapolations make any sense at all.
In the following we will alleviate this issue by considering, in the computation of the overall normalisation of the GCR flux, the supernova rate instead of the total number of star clusters.

\section{Spectrum of particles accelerated by clustered supernovae}\label{sec:spectraSN}

\subsection{First population: supernovae in loose clusters}
By definition, ``loose'' clusters do not sustain a collective outflow, as they are either too old (in which case there is not enough stellar power to form a collective WTS) or too extended (in which case the winds do not interact to produce a collective outflow). Individual stellar winds, as well as SNRs, deposit energy in the environment, which heats an expanding cavity filled by turbulent MHD outflows. Because the turbulent magnetic fields are diluted in this large cavity (sometimes referred to as a ``superbubble''), the average magnetic field probed by SNR shocks is not expected to exceed a few \textmu G and the maximum energy of the particles is estimated as \citep{lagage1983,hillas2005}:
\begin{align}
	E_{\rm max} \approx 0.1 \frac{Z B}{2~\text{\textmu G}} u_5^{1/3} \left(\frac{E_{\rm SN}}{10^{51}~\mathrm{erg/s}} \right)^{1/3} \left( \frac{n}{0.01 {{\rm cm}^{-3}}} \right)^{-\frac{1}{3}} \text{~PeV}
	\, ,
\end{align}
where $B$ is the average magnetic field, $n$ is the density within the superbubble cavity and $E_{\rm SN}$ is the explosion energy.
This computation assumes a constant shock velocity in the free expansion phase. A similar estimate is obtained when considering a decelerating shock, which is expected to be the case after the collision with the (individual) WTS of the progenitor star. Only the dependency on velocity changes, e.g. $E_{\rm max} \propto u_5^{2/3}$ for $R \propto t^{4/7}$, but this is not expected to affect the conclusions of this work. We set, for a loose cluster SN (LCSN) remnant shock expanding into a typical superbubble environment: 
\begin{equation}\label{Emaxlcsn}
    E_{\rm max,LCSN} \approx 0.1 Z u_5^{1/3} ~{\rm PeV} \, ,
\end{equation}
where the numerical factor can be taken as a parameter of the model.

Eventually we model the escaping differential energy spectrum for a given nucleus of charge $Z$ accelerated from a LCSN remnant shock as follows:
\begin{equation}
    S_{\rm LCSN} = \mathcal{A}_{\rm LC} \frac{E_{\rm SN} }{(p_0 c)^2} f_Z \left(\frac{p}{p_0}\right)^{-2-\Delta s} \exp \left( \frac{-pc}{E_{\rm max,LCSN}} \right) \, ,
\end{equation}
where $\mathcal{A}_{\rm LC}$ is a normalisation constant computed numerically such that $\int_{p_0} \dd p c \, E S_{\rm LCSN} \equiv f_Z E_{\rm SN}$\footnote{$S_{\rm LCSN}$ is related to the particle distribution function $f$ as $S_{\rm LCSN} = 4 \upi p^2 f$.}, with $p_0$ the injection momentum. $f_Z$ is therefore an acceleration efficiency which depends on the considered nucleus, in principle as a function of the composition of the medium and injection efficiency. In practice, we set the proton acceleration efficiency $f_1 \approx 0.1$ as an input parameter. The parametrization of the acceleration efficiencies $f_Z$ of heavier nuclei is then set {\it a posteriori} to match the composition measured at 1~TeV in the local GCR spectrum, from the fluxes tabulated in \citet{horandel2003}.

The parameter $\Delta s$ allows for spectral slope deviation from the canonical $p^{-4}$ scaling. This is motivated partly by gamma-ray observations displaying spectral indices typically around $2.1 - 2.5$ \citep[see e.g. Table 1 in][]{caprioli2011}. Although there is presently no convincing explanation of the discrepancy between these observations and the prediction of the standard diffusive shock acceleration theory, a number of arguments have been proposed, for example: the effects of nonlinear processes around the SNR shock \citep[e.g.][]{Kirk96,caprioli2012,vieu2021, Bell19}, magnetic-field obliquity \citep{Bell11,Xu22}, the weakening of the SNR shock after its collision with the wind termination shock of the progenitor star \citep[e.g.][]{das2022}, or the effect of the time-dependent particle leakage from the system \citep[e.g.][]{celli2019}. Within superbubbles, such a steepening might be even aggravated during the propagation of the particles in the turbulent environment, possibly up to very high energies \citep{bykov2001,ferrand2010}.

\subsection{Second population: supernovae in wind-blowing clusters}\label{sec:WBCSN}
By definition, ``wind-blowing clusters'' (WBC) are young and compact enough to power a collective wind. In this case, the energy deposited by stellar winds and SNRs is expected to excite strong turbulence within the cluster. The turbulent energy decreases in the free-wind region which extends beyond the boundary of the cluster (e.g. as $1/r^2$). Therefore, the magnetic field induced within the cluster stays confined within the cluster and can in principle reach large values, up to hundreds of \textmu G \citep{vieu2022Emax}. This phenomenological argument is supported by recent 3D MHD simulations of the inner core of a WBC \citep{badmaev2022}, which showed that even without resolving the small-scale turbulence, the kinetic stellar energy is efficiently converted to magnetic energy {\it via} the Cranfill effect, thereby creating magnetic fields up to several hundreds of \textmu G in filamentary structures. Although these simulations do not probe the effect of SN explosions nor the properties of the collective outflow beyond the boundaries of the cluster, they do showcase that efficient magnetic field amplification is to be expected in and around WBCs. A SNR shock expanding into these large magnetic fields could accelerate protons up to a maximum energy which is limited by the available time.
The acceleration rate, assuming diffusion at the Bohm limit, is $\dd p/\dd t \approx 3 q B/(\alpha c) \dot{R}^2$, where $\alpha \approx 9$ if one assumes that the upstream magnetic field $B$ is compressed by a factor $\sqrt{11}$ across the shock. Because the magnetic field decreases in the collective wind beyond the boundary of the cluster ($B\propto 1/r$), the acceleration rate is higher around a SNR shock expanding inside the cluster than around a SNR shock launched close to the edge and expanding in the cluster wind. However the fate of SNRs expanding deep inside the cluster is unclear. Collisions with the strong winds of the neighbouring massive stars are expected to hinder, at least partly, the propagation of the shock, eventually converting part of the SNR shock pressure into MHD turbulence. Although a system of converging shocks might be in itself an efficient accelerator of very-high energy CRs \citep{bykov2015}, it is yet unclear if the increase of the maximum energy compared to a single shock is a generic outcome of this class of systems \citep{vieu2020}. In the following, we therefore assume that such collisions between shocks in compact clusters do not leave noticeable spectral features at ultra-high energies.

If a SN explodes deep within the cluster, its remnant shock typically reaches the edge of the cluster at the time of the transition to the Sedov-Taylor phase, which implies that particles at the highest energy are not further accelerated in the cluster wind, but on the contrary suffer adiabatic losses when advected from the cluster edge ($\sim 1$~pc) to the cluster WTS ($\sim 10-20$~pc). This shifts the spectrum back to lower energies, such that, in the end, the maximum CR energy of the particles escaping the spectrum is typically higher if the SN shock is launched close to the cluster edge and expands mostly in the collective wind. As a conservative approach, we only consider the latter in the following. The fraction of SN which produce shocks expanding far in the wind can be estimated as follows. Assuming that $N_*$ massive stars are homogeneously distributed in the cluster, the distance between them reads $L = (2/3)^{1/3} R_{\rm c} N_*^{-1/3}$. About $2\pi R_{\rm c}/L$ stars lie at the edge of the cluster, which represents a fraction $\eta_{\rm SN} \approx 7 N_*^{-2/3} \approx 0.3 - 0.5$ of all SN progenitors. The parameter $\eta_{\rm SN}$ can be taken as an efficiency which globally accounts for the aforementioned uncertainties about the fate of SN shocks launched within compact star clusters.

Let us now estimate the maximum energy of particles accelerated around a SNR shock which expands in the collective wind. Assuming a $1/r$ magnetic field profile in this region, we get, from the acceleration rate:
\begin{equation}
    E_{\rm max} \approx 3 q B_{\rm c} R_{\rm c} /(\alpha c) u_{\rm SN} \ln \left(R_{\rm max}/R_{\rm c} \right)\, ,
\end{equation}
where $B_c$ is the value of the magnetic field at the edge of the cluster, $R_{\rm c}$ is the radius of the cluster, $u_{\rm SN}$ is the initial SN shock velocity, $R_{\rm max}$ is either the distance reached at the end of the free expansion phase, or the radius of the WTS. For fast SN shocks (30~000~km/s) launched close to the edge of the cluster, one can compute\footnote{We assume that the energy released in each SNR is $\approx 10^{51}$\,ergs}
$R_{\rm max}/R_{\rm c} \approx 5$,
while for slow SN shocks (5000~km/s), the shock collides with the WTS while being still in the free expansion phase and $R_{\rm max}/R_{\rm c} \sim 10$.
Neglecting this small logarithmic correction, we obtain:
\begin{equation}
    E_{\rm max} \approx 3 Z \left(\frac{B_{\rm c}}{100~\text{\textmu G}}\right) \left(\frac{R_{\rm c}}{5~\text{pc}}\right) u_5 ~\mathrm{PeV} \, ,
\end{equation}
where $u_5$ is again the initial shock velocity in units of 5000~km/s. Taking into account the deceleration of the shock in the wind profile \citep[\(u \propto t^{-1/8}\),][]{finke2012,gaggero2018} leads to a similar estimate, as detailed in \citet{vieu2022Emax}. We can generically write:
\begin{equation}\label{timelimitSNRcompactcluster2}
    E_{\rm max,WBCSN} \approx Z u_5 f_{\rm c}(R_{\rm c},N_*,n_{\rm c},\eta_{\rm T}) ~\text{PeV} \, ,
\end{equation}
where $f_{\rm c} \sim O(1)$ is a function of the cluster properties (cluster radius $R_{\rm c}$, number of massive stars $N_*$, density within the cluster $n_{\rm c}$, and the fraction of the stellar mechanical power injected into the MHD turbulence cascade $\eta_{\rm T}$). As we aim to obtain the average contribution from all clusters in the Galaxy, and because the SN shock velocity $u_5$ is not expected to correlate with the properties of the cluster, the function $f_{\rm c}$ can be estimated on average. It is furthermore not meaningful to compute it in detail as most of the cluster parameters are uncertain, both from the observational and theoretical perspectives. In particular, it is difficult to constrain the turbulence generation efficiency $\eta_{\rm T}$ or the density outside of the superbubble cavity. In the following we take the following estimate for an average galactic cluster (assuming $\eta_{\rm T} = 10\%$):
\begin{equation}\label{Emaxwbcsn}
    E_{\rm max,WBCSN} \approx 4 Z u_5~\text{PeV} \, ,
\end{equation}
which can be taken as an input parameter of the model, justified by reasonable values of the average cluster properties in Eq.~\ref{timelimitSNRcompactcluster2} \citep[details can be found in][]{vieu2022Emax}.

In the scenario considered here, the magnetic field is not amplified via streaming instabilities but rather from a deposition of energy at the stellar scale within the boundaries of the cluster. It is assumed that, if necessary, CRs still sustain the scattering fields required to ensure Bohm diffusion.
These SNR shocks are therefore likely to produce hard spectra of high-energy particles which we assume to match the standard $p^{-4}$ power-law achieved around a strong shock, such that we can write:
\begin{equation}
    S_{\rm WBCSN} = \mathcal{A}_{\rm WBC} \frac{E_{\rm SN}}{2} f_Z  (pc)^{-2} \exp \left( \frac{-pc}{E_{\rm max,WBCSN}} \right) \, ,
\end{equation}
where $\mathcal{A}_{\rm WBC}$ is a normalisation constant set such that $\int_{p_0} \dd p c \, E S_{\rm WBCSN} \equiv f_Z E_{\rm SN}/2$. Only about half of the shock is launched in the collective wind region if the supernova explodes close to the cluster edge, hence the additional factor of 1/2 in the normalisation. The acceleration efficiency $f_Z$ is assumed to be equal to that of the LCSN remnant shocks, even though the composition of the wind blown by a WBC differs from that of a LC, due to the presence of WR stars. Figure~3 in \citet{tatischeff2021} suggests that the contribution from the latter does not make a dramatic difference and thus will not affect our results substantially. A detailed study of the WBC plasma composition relative to that of LCs, taking into account the material launched by the stellar population, the material deposited by expanding SNR, and the mass-loading from the ISM, is beyond the scope of this paper.

\section{Galactic CR spectrum from massive star clusters}\label{sec:galspectrum}
In this section we consider the entire population of galactic MSCs as analysed in Section~\ref{sec:MSCpopulation}, in order to compute their contribution to the flux of particles detected near the Earth.

\subsection{Spectrum of particles accelerated by stellar winds}\label{sec:stellarwinds}
In the previous section we computed the contribution of the clustered SNR shocks to the GCR population. However, massive stars blow powerful winds during their lives, which create wind termination shocks (WTS), and it was early realised that these could be favourable sites of particle acceleration \citep[e.g.][]{casse1980,cesarsky1983}. In this section we consider the contribution of stellar winds in the model.

We first estimate the average mechanical power per massive star in young and evolved clusters, which is critical to infer the normalisation of the wind components. Stellar evolution models can be used to follow the evolution of a population of clustered massive stars from the formation of the cluster (usually one assumes a coeval population of stars) to the explosion of the last SN, which by our definition marks the end of the MSC life. In the young phase, the mechanical power of MSCs is dominated by the outflows of early O and WR stars, while in the evolved phase, where by definition there is no remaining WR stars, the power drops significantly. Assuming a Salpeter IMF and using the fits of stellar evolution models provided by \citet{seo2018} for main sequence O and early B stars, we compute an average mechanical power per massive star of $5 \times 10^{32}$~erg/s for evolved clusters. By comparison, each clustered supernova provides $10^{51}$~erg, which, averaged over a cluster lifetime of 40~Myr, represents a power of $8 \times 10^{35}$~erg/s. The mechanical power input of stellar winds embedded in evolved clusters can therefore be neglected.

\begin{figure*}
          \centering
              \includegraphics[width=\linewidth]{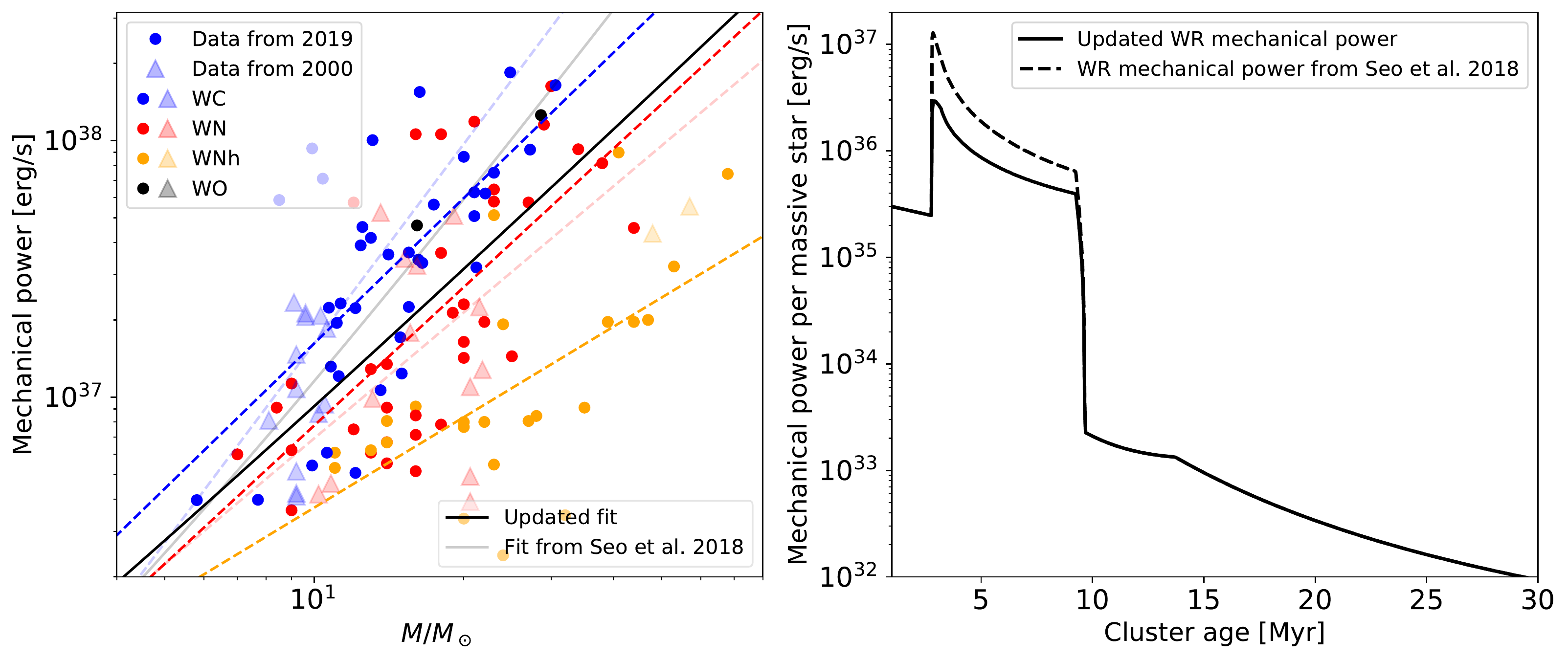}
   \caption{Left: Mechanical power of WR stars as a function of their mass and type. The blue, red and yellow dashed lines show the linear fit to the individual types whilst the solid black line shows the average contribution assuming an equal number of WC, WN and WNh stars. These fits take into account the updated data from \citet{sander2019,hamann2019} (points), to be compared with the transparent lines which fit the oldest data from \citet{nugis2000} (triangles). Right: Evolution of the mechanical power per massive star as function of the age of an average massive star cluster, assuming a coeval stellar population initially distributed according to a Salpeter IMF.}
   \label{fig:starpower}
 \end{figure*}

To compute the mechanical power of massive stars embedded in young clusters, it is critical to know the contribution from the Wolf-Rayet stars. \citet{seo2018} estimated this quantity by fitting observational data from \citet{nugis2000} (shown by the triangles in the left panel of Figure~\ref{fig:starpower}). However the lack of observations of WR stars more massive than $25 M_\odot$ and the dispersion of the data make it difficult to extrapolate this analysis to high stellar masses. In 2019, the population of galactic WN and WC stars was revisited and complemented in light of the Gaia DR2 \citep{sander2019,hamann2019}. The updated list of single WR stars contains WC stars up to $30 M_\odot$, WN stars up to $45 M_\odot$ and WNh stars up to $70 M_\odot$, which makes the extrapolation of the data more meaningful. As shown in the left panel of Figure~\ref{fig:starpower}, the updated relation between mechanical power and star mass (solid black line) deviates significantly from that obtained by fitting the data of \citet{nugis2000} (transparent black line), with a much shallower slope, which implies that high mass WR stars are less powerful than previously thought, by up to one order of magnitude. Including these revised relations in a time-dependent cluster model, we compute, for young clusters following a Salpeter IMF, an average mechanical power per massive star of $4 \times 10^{35}$~erg/s if the fit is performed ignoring the WR spectral type. The average increases to $7 \times 10^{35}$~erg/s if we fit the three WR spectral types (WC, WN, WNh) independently and then assume that WR stars are equally distributed between these three types\footnote{This is likely providing an upper bound as it does not account for the fact that the WC type is actually a stellar evolution phase subsequent to the WN phase, hence the average mass of WC stars is expected to be significantly smaller than that of WN stars, as suggested by the observational data (Figure~\ref{fig:starpower}).}. In the following, we will assume an average mechanical power per massive star embedded in a young cluster of $5 \times 10^{35}$~erg/s\footnote{In comparison, the relations found by fitting the outdated data from \citet{nugis2000} provide a significantly higher average power of $1.7 \times 10^{36}$~erg/s.}. Integrated over several Myr, this power provides an energy input comparable to that of a supernova, hence the contribution of the young cluster stellar winds to the galactic CRs cannot be disregarded.

In order to compute the normalisation of the wind component, we also need to estimate the total number of massive stars in the Galaxy. This can be done in two independent ways, either considering the birth or death rate. First, the average star formation rate of $2 M_\odot$/yr in the Milky Way is expected to generate a population of about 450k massive stars, assuming that massive stars are distributed according to a Salpeter initial mass function and live on average for 17~Myr. Second, to sustain a rate of core-collapse supernova of 2 per century, there must be at least 340k massive stars of average lifetime 17~Myr. As a compromise we assume the total number of massive stars to be 400k. Then, knowing the fraction of WBC as determined in Section~\ref{sec:MSCpopulation}, and assuming that about $80-90$\% of massive stars are clustered \citep{higdon2005,lingenfelter2018}, we can infer the total number of stars embedded in WBCs. Finally, one estimates the total wind power of galactic WBCs as:
\begin{equation}
    P_{\rm w,WBC} = f_{\rm WBC} f_{\rm c} N_{\rm *,Gal} \bar{P}_{\rm w} \, ,
\end{equation}
where $f_{\rm WBC} \approx 15\%$ is the fraction of WBCs in the galactic MSC population (Section~\ref{sec:MSCpopulation}), $f_{\rm c} \approx 80-90 \%$ is the probability that a massive star is embedded in a MSC, $N_{\rm *,Gal} \approx 400k$ is the total number of massive stars in the Galaxy, and $\bar{P}_{\rm w} \approx 5 \times 10^{35}$~erg/s is the average mechanical power per massive star in a young cluster as discussed above.

WBCs by definition blow a collective wind which terminates at a distance of order 10~pc from the centre of the cluster. Particle acceleration is expected to take place at the wind termination shock up to about 1~PV, for favourable conditions  \citep{gupta2020,morlino2021,vieu2022Emax}. We therefore model the spectrum of accelerated particles as follows:
\begin{equation}\label{phiwind}
    \phi_{\rm w} = \mathcal{A}_{\rm w} f_Z  (p c)^{-2} \exp \left(\frac{-pc}{E_{\rm max,WTS}} \right)\, ,
\end{equation}
where the normalisation constant $\mathcal{A}_{\rm w}$ is set such that $\int_{p_0} \dd p c \, E \phi_{\rm w} \equiv f_Z P_{\rm w,WBC}$, and the acceleration efficiency $f_Z$ is assumed to be the same as for the case of strong SNR shocks. Gamma-ray observations of star clusters typically display spectra which are slightly steeper than $E^{-2}$ \citep[e.g.][]{gammaWd2_yang2018,gammaCygnus_HAWC2021,HESSWd12022}, however the hard wind component calculated from Eq.~\ref{phiwind} will be already found in the following to provide a subdominant contribution to the GCR flux -- which can be anticipated as the integrated power of stellar winds is about half that of SN explosions --, such that the shape of the spectrum given by Eq.~\ref{phiwind} is not expected to impact our results.

\subsection{Escaping spectra from a distribution of sources}
The maximum energy achieved in SNR shocks depends on the initial velocity of the shock (Eqs.~\ref{Emaxlcsn} and \ref{Emaxwbcsn}). We therefore need to introduce a distribution of velocities denoted $\xi$, and defined as:
\begin{equation}
    \dd f_{\rm SN} = \xi(u) \dd u \, ,
\end{equation}
such that $f_{\rm SN} = \int \xi(u) \dd u$ is the total rate of core-collapse SN (CCSN) in our Galaxy. This distribution is difficult to constrain. As discussed in \citet{Sveshnikova2003}, it is challenging to interpret the distribution of SN absolute-magnitudes derived from observations, and it is further difficult to constrain the initial velocity of the SNR shock. In the following we assume that SN shock velocities range between 3000~km/s and 30~000~km/s (e.g. SN1987A), with a power-law distribution, e.g. $\xi(u)\propto 1/u$. The sensitivity of our results on the shape of the distribution $\xi$ is discussed in the end of Section~\ref{sec:allparticles}.

Finally, we gather the two SN components together with the wind component to obtain the total flux released by MSCs in our Galaxy:
\begin{multline}\label{inj_S_protons}
    \phi_{\rm c}(E,Z) = (1-f_{\rm WBC}) f_{\rm c} \int \dd u \, \xi(u) S_{\rm LCSN} \\
    + f_{\rm WBC} f_{\rm c} \eta_{\rm SN} \int \dd u \, \xi(u) S_{\rm WCSN} + \phi_{\rm w} \, ,
\end{multline}
where $f_{\rm WBC} \approx 15\%$ is the fraction of WBC in the galactic MSC population obtained in Section~\ref{sec:MSCpopulation}, $f_{\rm c} \approx 80\%$ is the clustering fraction of massive stars and, as discussed in Section~\ref{sec:WBCSN}, $\eta_{\rm SN}$ accounts for the fact that not all WBCSN remnant shocks are able to accelerate particles up to the maximum energy given by Eq.~\ref{Emaxwbcsn}, especially if their progenitor explodes deep inside the cluster.

\subsection{Propagation}
Eq.~\ref{inj_S_protons} provides the spectrum of a given element of atomic number $Z$ released in the Galaxy. Diffusive propagation in the galactic magnetic fields is then taken into account using a one-zone model, which allows one to write the spectral energy distribution of relativistic particles (per unit surface, per unit solid angle, per unit time) as:
\begin{equation}\label{dNdE_Z}
    \left. \frac{\dd N}{\dd E} \right|_{Z} = \frac{H_{\rm disk}^2}{V_{\rm disk}} \frac{c}{4\pi D_{\rm Gal}(E,Z)} \phi_{\rm c}(E,Z) \, ,
\end{equation}
where $H_{\rm disk}$ and $V_{\rm disk}$ are respectively the height and volume of the Galactic disk, and $D_{\rm Gal}$ is the galactic diffusion coefficient parametrized as follows:
\begin{equation}\label{diffusioncoeff}
    D_{\rm Gal}(E,Z) = D_0 \left(\frac{E}{Z} \right)^\delta \, .
\end{equation}
The value of the galactic diffusion coefficient at 1~GV only affects the global normalisation of the spectrum and can be adjusted together with the acceleration efficiency. In the following we assume $D_0 = 10^{28}$~cm$^2$/s. The parametrization given by Eq.~\ref{diffusioncoeff} is only valid up to the rigidity for which the particle mean free path $D/c$ equals the scale height of the Galactic disk. Assuming the latter to be 500~pc gives a break rigidity of $(5 \times 10^3)^{1/\delta}$~GV, which, for consistency, must be larger than the maximum rigidity achieved around the fastest WBCSN remnant shocks. This implies that $\delta$ must be smaller than 0.5 (smaller than 0.4 if we assume a disk height of 100~pc). In the following, the best fit value will be found to be $\delta = 0.35$, therefore meeting this requirement.

\begin{table*}
    \centering
    \begin{tabular}{lcc}
    \bf Parameter & \bf Notation & \bf Value \\ \hline
         Minimum initial velocity of a SN shock & $u_{\rm min}$ & 3000~km/s \\
         Maximum initial velocity of a SN shock & $u_{\rm max}$ & 30000~km/s \\
         Scaling of SN shocks velocity distribution & $\xi(u)$ & $\propto u^{-1}$ \\
         Fraction of massive stars which are clustered & $f_{\rm c}$ & 80\%  \\
         Rate of CCSN & $f_{\rm SN}$ & $2/$century \\
         Average energy of a SN explosion & $E_{\rm SN}$ & $10^{51}$~erg
         \\
         \hline
         \textbf{$^*$Fraction of WBC in the MSC population} & $f_{\rm WBC}$ & \bf 16\% \\
         \textbf{$^*$Probability to launch a fast shock in the collective WBC wind} & $\eta_{\rm SN}$ & \bf 0.5 \\
         \hline
         Wind mechanical power per young massive star & $\bar{P}_{\rm w}$ & $5 \times 10^{35}$~erg/s \\
         Number of massive stars in the Galaxy & $N_{*,\rm Gal}$ & 400k \\
         Maximum energy achieved at WTS & $E_{\rm max,\rm WTS}$ & 1~PeV \\
         \hline
         \textbf{$^*$Proton injection efficiency} & $f_1$ & \bf 0.18 \\
         Composition at 1~TeV & $f_Z$ & \citet{horandel2003} \\
         \textbf{$^*$Spectral index deviation between LCSN and WBCSN spectra} & $\Delta s$ & \bf 0.17 \\
         CR injection momentum & $p_0$ & 10~MeV/c \\
         Maximum energy of protons from a 5000~km/s LCSN shock & - & 0.1~PeV \\ 
         Maximum energy of protons from a 5000~km/s WBCSN shock & - & 4~PeV \\ 
         \hline
         \textbf{$^*$Galactic diffusion coefficient scaling} & $\delta$ & \bf 0.35 \\
         Galactic diffusion coefficient at 1~GV & $D_0$ & $10^{28}$~cm$^2$/s \\
         Height of the Galactic disk & $H_{\rm disk}$ & 0.5~kpc \\
         Volume of the Galactic disk & $V_{\rm disk}$ & 400~kpc$^3$ \\
    \end{tabular}
    \caption{Exhaustive list of input parameters. Parameters marked with a star are ``fitting'' parameters, which are allowed to vary within reasonable bounds (see text).}
    \label{tab:parameters}
\end{table*}

\subsection{All-particle spectrum}\label{sec:allparticles}
We compute the all particle spectrum by summing Eq.~\ref{dNdE_Z} up to $Z=31$. While the normalisation of the proton spectrum is set by the acceleration efficiency parameter, the contribution from each nucleus is then normalised with respect to that of protons to ensure that we retrieve the elemental abundances at 1~TeV given in \citet{horandel2003}.
The resulting spectrum depends on a number of parameters, which are listed in Table~\ref{tab:parameters}. Highlighted in bold are parameters which we allow to vary in order to fit the data. The other parameters are fixed to the values indicated in the table.

Among the fitting parameters, the proton injection efficiency sets the overall normalisation of the spectrum. The slope is affected by the galactic diffusion coefficient scaling $\delta$, while the relative steepness between the LCSN and WBCSN components (i.e. the shape of the ``knee'') can be adjusted by varying the spectral index deviation $\Delta s$.
The fraction of WBC in the MSC population sets the relative normalisation of the two dominant LCSN and WBCSN components, while the parameter $\eta_{SN}$, which encodes the uncertainties about the fate of SNRs expanding deep inside WBCs, impacts the relative normalisation of the wind component. We vary these five parameters in order to fit the all-particle spectrum of CRs. It should be noted that these parameters are tightly constrained theoretically. For instance, the galactic diffusion coefficient index is not expected to be smaller than $1/3$ (which corresponds to a Kolmogorov turbulence regime). However, a higher diffusion coefficient index leads to a steeper spectrum and therefore requires an increase of the normalisation in order to match the measured flux. This can only be achieved by raising the proton injection efficiency above 20\%, which becomes higher than what is suggested by theory. In the end, the five parameter values displayed in Table~\ref{tab:parameters} are the result of a compromise between the normalisation of the components and the shape of the spectrum.

\begin{figure*}
          \centering
              \includegraphics[width=\linewidth]{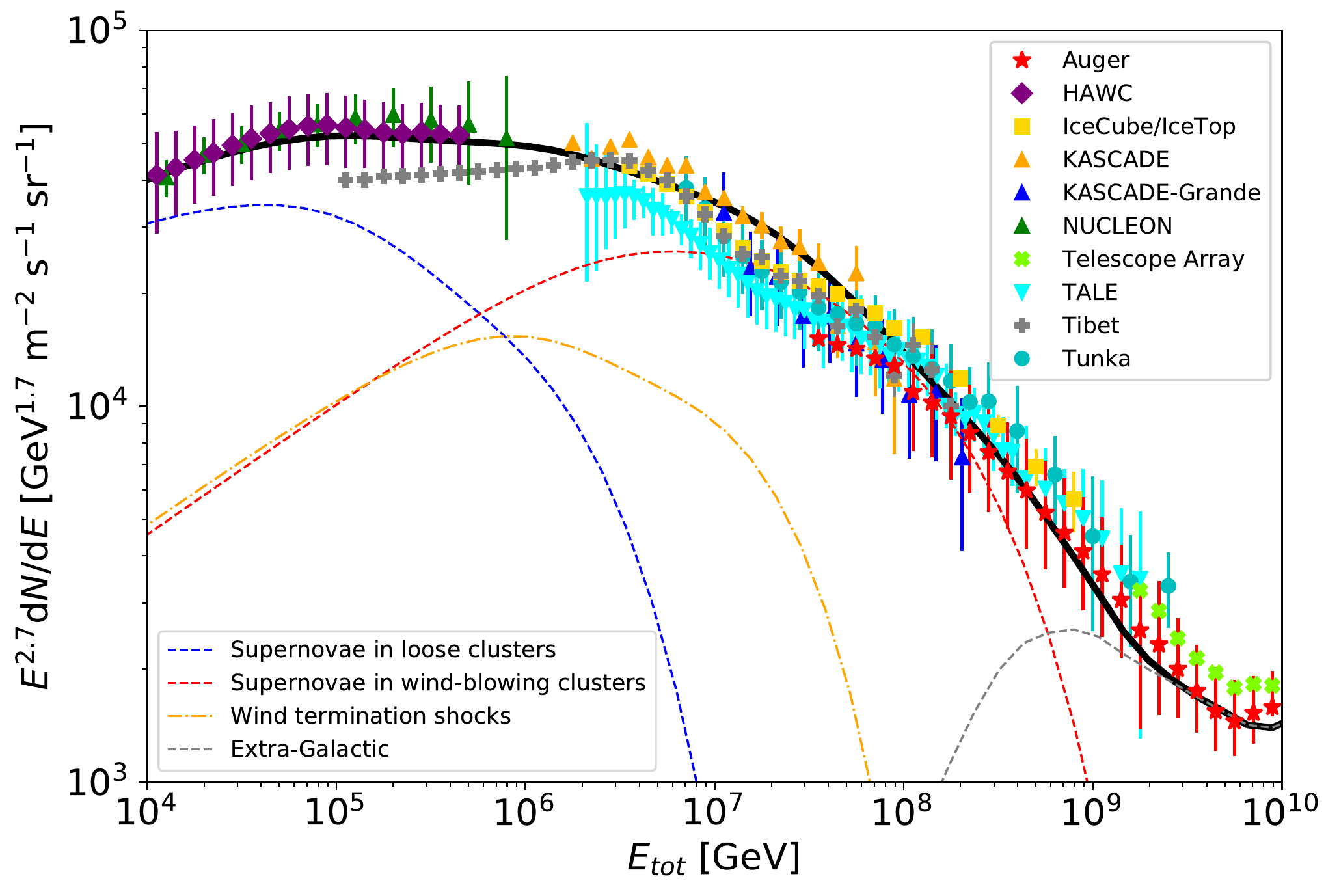}
   \caption[]{All-particle spectrum obtained from our computation, compared with recent data, partly extracted from the compilation by C.~Evoli (\url{https://github.com/carmeloevoli/The_CR_Spectrum}): Auger \citep{DataAP_Auger,dataAuger2021}, HAWC \citep{DataAP_Hawc}, IceCube/IceTop \citep{DataAP_Icecube}, KASCADE \citep{DataAP_Kascade}, NUCLEON \citep{DataAP_Nucleon}, Telescope Array \citep{DataAP_TA}, Tibet \citep{DataAP_Tibet}, Tunka \citep{DataAP_Tunka}.}
   \label{fig:result_allpart}
 \end{figure*}

The all-particle spectrum obtained subsequently to this fitting procedure is shown in Figure~\ref{fig:result_allpart}. It displays a good agreement with the measurements, given the spread of the data. In particular, our calculation results in an excellent match with the HAWC, Nucleon, Kascade, Tunka and Auger data, from 10~TeV to a few hundreds of PeV. The galactic contribution at UHE, dominated by WBCSN (red dashed line), succeeds in reproducing the data up to 200~PeV, and starts to undershoot the data beyond 300~PeV.
This demonstrates that supernovae exploding within MSCs can account for the UHE GCR spectrum with reasonable choice of the parameters. In particular, we show that a few fast SN shocks (2/millenium) can alone account for the particles seen beyond 10~PeV, up to hundreds of PeV.

As mentioned in the introduction, composition measurements point to a transition from GCRs to extragalactic CRs taking place at a few hundreds of PeV, with data suggesting the mean atomic mass peaks at $100$~PeV and then decreases up to the ankle \citep[e.g.][]{supanitsky2022}. In order to probe how a sub-ankle extragalactic component complements our scenario for the origin of GCRs, we added the contribution modelled in \citet{unger2015} (considering the photo-disintegration of nuclei close to the extragalactic sources). To this aim, we reproduce the ``UFA'' component calculated in \citet{thoudam2016}, which includes propagation effects. It is shown by the grey dashed line in Figure~\ref{fig:result_allpart}. This showcases that a sub-ankle extragalactic component can indeed compensate the fall-off of the galactic contribution, leading to a very good fit of the Auger data up to the ankle. Similar conclusions are obtained using the calculations of \citet{globus2015} or the model by \citet{kotera2008}.

\begin{figure}
          \centering
              \includegraphics[width=\linewidth]{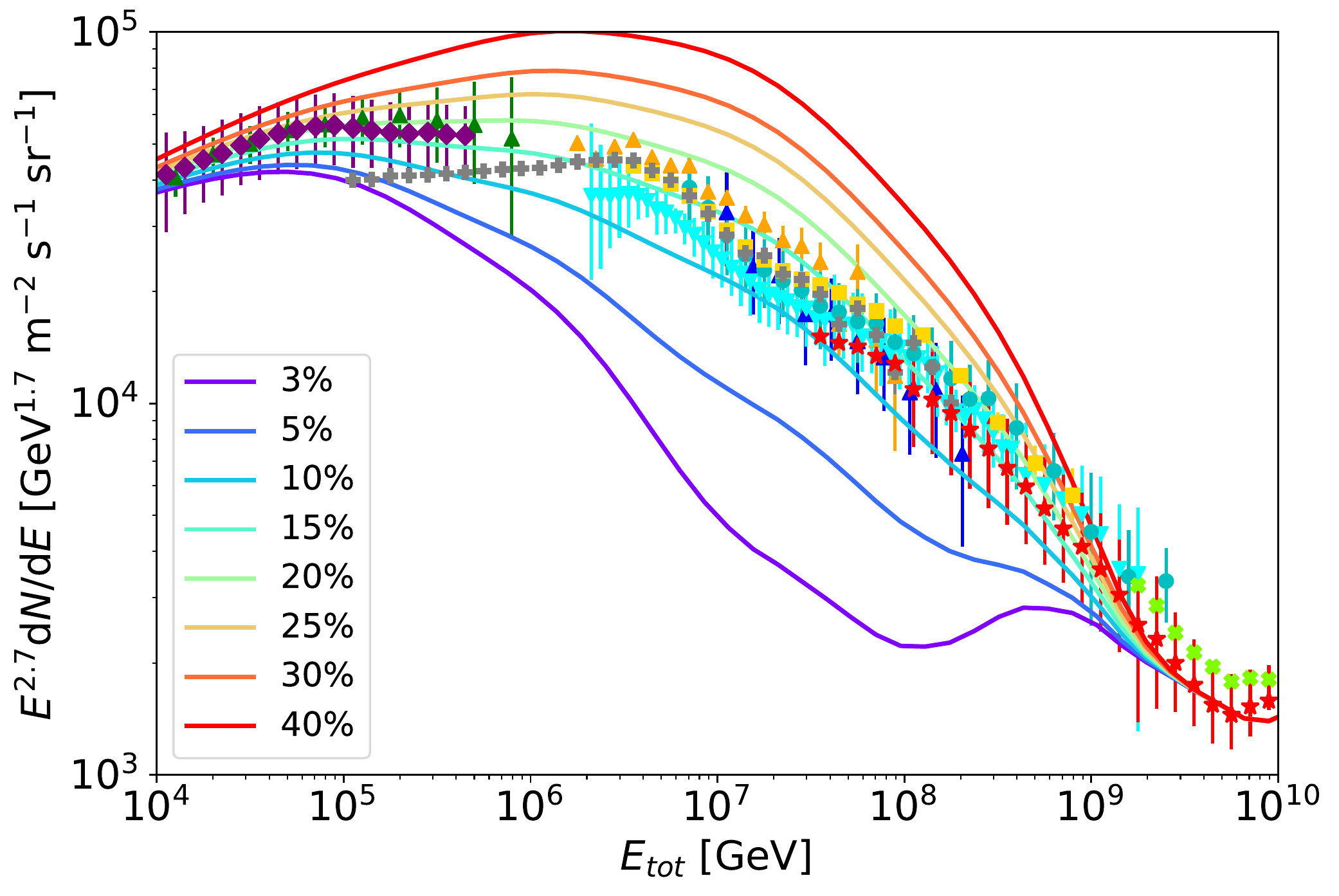}
   \caption{All-particle spectrum obtained by varying the fraction of wind-blowing clusters $f_{\rm WBC}$ with values indicated in the legend. The other parameters are kept as listed in Table~\ref{tab:parameters}.}
   \label{fig:result_finetuning}
 \end{figure}

Any model trying to reproduce, with two components, a concave transition between two featureless power-laws must require some level of tuning in order to avoid unexpected bumps or drops \citep{parizot2014}. We show in Figure~\ref{fig:result_finetuning} that the fraction of WBCs, which sets the normalisation of the second, subdominant, component in our model, must indeed be within 5\% to 25\% in order to reproduce a knee shape. A fraction smaller than 5\% produces a drop around 100~PeV, while a fraction larger than 25\% produces a bump around 1~PeV. Without prior knowledge on the nature of this second component, it might seem to be a coincidence to fall in this narrow range. However, the knowledge that this component comes from a subset of the galactic MSCs already sets a strong constraint on its normalisation factor, as it implies that it must be smaller than 1 and, in fact, most likely smaller than 50\% since we know that the majority of clusters are not blowing a collective wind. More detailed investigation on the star cluster population in the Milky Way, as done in Section~\ref{sec:MSCpopulation}, show that the fraction of WBCs falls exactly in the required range (about 16\%, keeping in mind the uncertainties discussed in Section~\ref{sec:MSCpopulation}) and no posterior adjustment on this parameter is needed. 

\begin{figure}
          \centering
              \includegraphics[width=\linewidth]{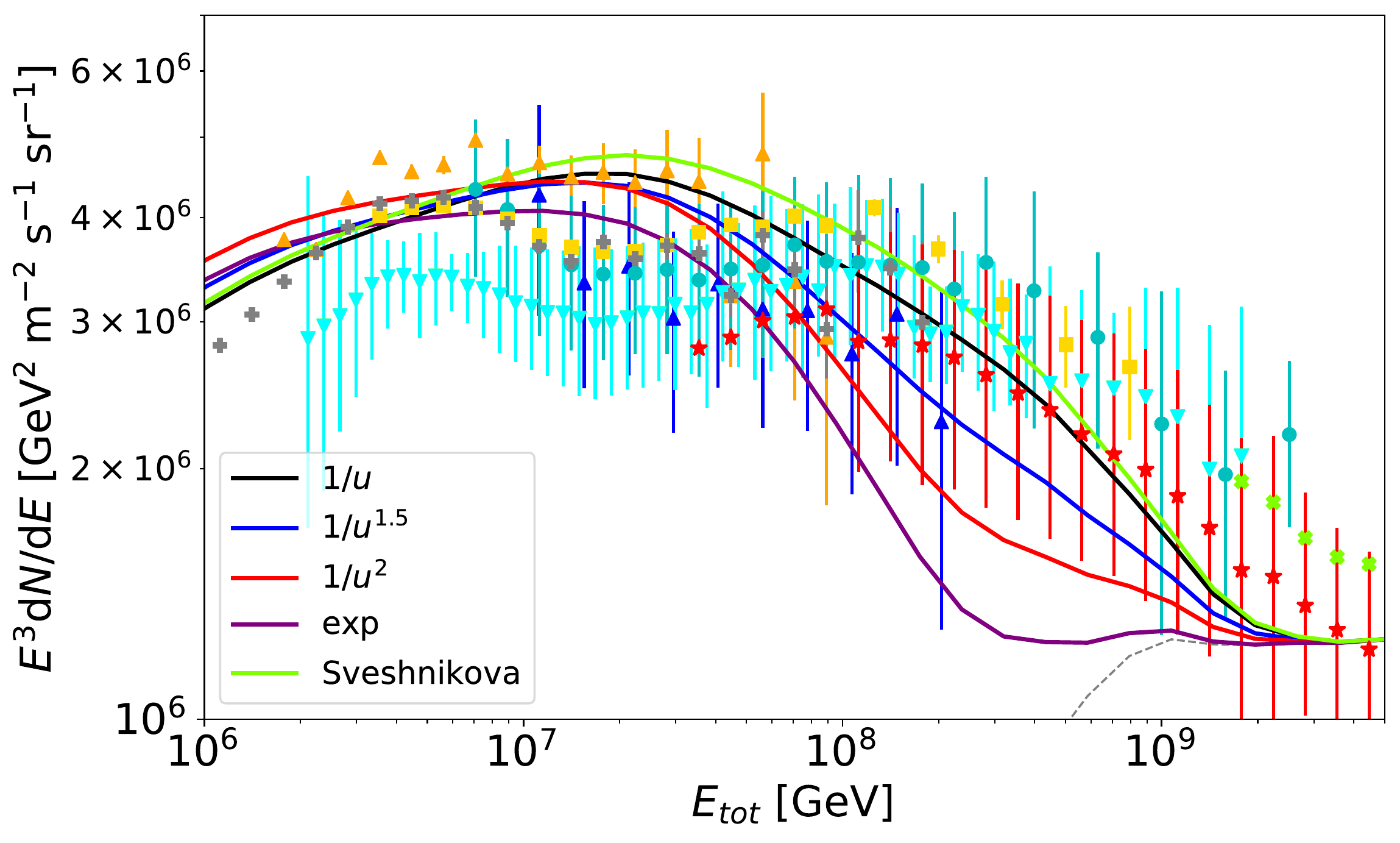}
   \caption{All-particle spectrum obtained from various assumptions on the distribution of SN shock velocities $\xi$: three different power-laws, an exponential cut-off at $u=3000$~km/s, and a model inspired by the work of \citet{Sveshnikova2003} where the explosion energy follows as well a power-law distribution and the normalisation of the CR spectra is set accordingly (see text). All parameters are kept as listed in Table~\ref{tab:parameters}, except for the latter case where $f_1 = 0.14$.}
   \label{fig:result_varyphi}
 \end{figure}

Our model relies on the assumption that the distribution of SN shock velocities can be described by an extended tail. With a distribution of SN shock velocities scaling as $1/u$, one expects two SN per millennium launching a shock with velocity 30 000~km/s. This is obviously required since the maximum proton energy for a standard 5000~km/s shock expanding close to the edge of a WBC is 4~PeV, leading to a maximum iron energy of 100~PeV, which would not be enough to reproduce the all-particle spectrum up to the ankle, even considering a sub-ankle extragalactic component. We show in Figure~\ref{fig:result_varyphi} the results which would have been obtained assuming different scalings of the distribution of SN shock initial velocities $\xi(u)$. Only a distribution $\xi(u)=1/u$ is able to reproduce the Auger data up to 1~EeV. A $1/u^2$ scaling fails to match the systematics errors and, as expected, an exponentially decreasing distribution falls short of the data beyond 100~PeV. This shows that the $1/u$ distribution of SNR shock velocities is a critical ingredient of our model, together with the maximum proton energy achieved around standard SNR shocks expanding close to WBCs.

In contrast, the analysis done by \citet{Sveshnikova2003} suggests a steeper distribution tail, scaling as $\sim u^{-3}$. This is however inferred from an energy distribution tail extending up to about $10^{53}$~erg (``hypernovae'') under the hypothesis of constant ejected mass. The latter is questionable, and one can find a hint in recent numerical simulations that the ejected mass is anti-correlated with the explosion energy \citep{ebinger2020}. For instance, SN1987A had an estimated explosion energy around $10^{51}$~erg but launched a shock of velocity beyond 30~000~km/s \citep{blinnikov2000}. Nevertheless, we can incorporate the ``Sveshnikova'' distribution in our model, assuming $\xi(u) \propto u^{-3}$, and, for simplicity, $E_{\rm SN} = u_5^2 \times 10^{51}$~erg. The result is shown by the light green curve in Figure~\ref{fig:result_varyphi}. Due to the assumption of constant ejected mass, the fastest SN shocks are also the most energetic. This makes their contribution to the CR spectrum larger, which compensates for their scarcity. Incidentally, to obtain the result shown in Figure~\ref{fig:result_varyphi} in this specific case, it was necessary to reduce the proton injection efficiency to 14\% in order to avoid overshooting the data. This is the only parameter which differs compared to the other curves.

In the end, it is unclear which model of SN distribution is the closest to reality, which is why, in the present paper, we agnostically chose to stick to a constant explosion energy of $10^{51}$~erg and to set fiducial velocity distributions.

\subsection{Proton spectrum}
\begin{figure}
          \centering
              \includegraphics[width=\linewidth]{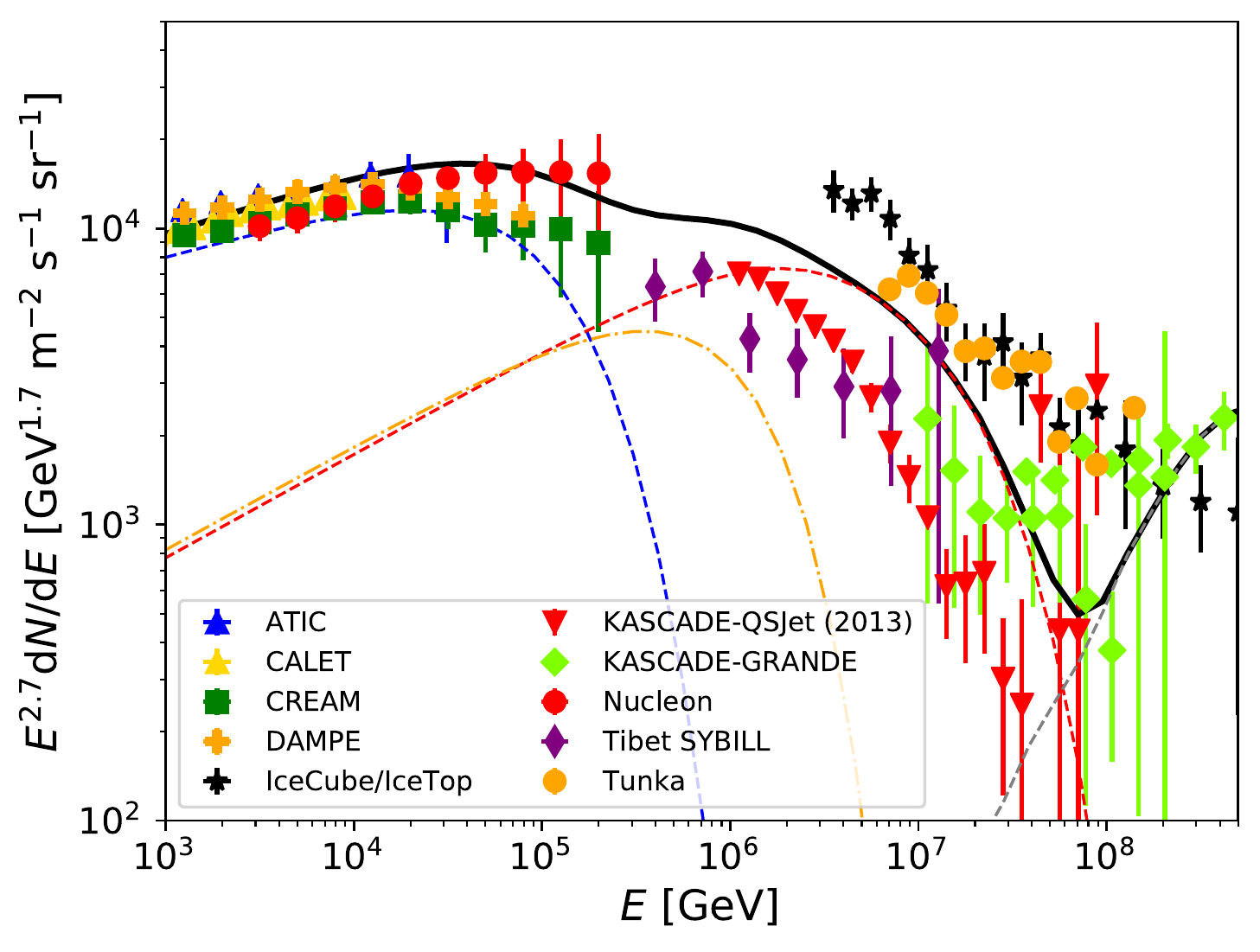}
   \caption{Proton spectrum obtained from our computation, compared with recent data \citep{Data_p_ATIC2009,Data_p_CALET2019,Data_p_CREAM2017,Data_p_DAMPE2019,DataAP_Icecube,Data_p_Kascade2005,Data_p_KascG2013,DataAP_Nucleon,Data_p_Tibet2006,Data_p_TUNKA2013}.}
   \label{fig:result_protons}
 \end{figure}

Having computed the all-particle spectrum, we can extract the spectrum of the protons and compare it with available data. The result is shown in Figure~\ref{fig:result_protons}. Interpreting the match between our model and the data is less straightforward in the case of the protons. Indeed, the discrepancy between experiments, mostly due to the high uncertainties in the modelling of hadronic air showers, does not set tight constraints beyond 1~PeV. We nevertheless find good agreement below 1~PeV, where our model meets most of the data within error bars. It does seem that the WBCSN contribution tends to overshoot the Kascade and Tibet measurements above 1~PeV, however this is actually expected, as WBCSN accelerate mostly material blown by WR stars, whose composition contains less hydrogen than the average superbubble composition around loose clusters \citep[about 80\%, as suggested by Figure~3 in][]{tatischeff2021}.

\subsection{Composition}
\begin{figure}
          \centering
              \includegraphics[width=\linewidth]{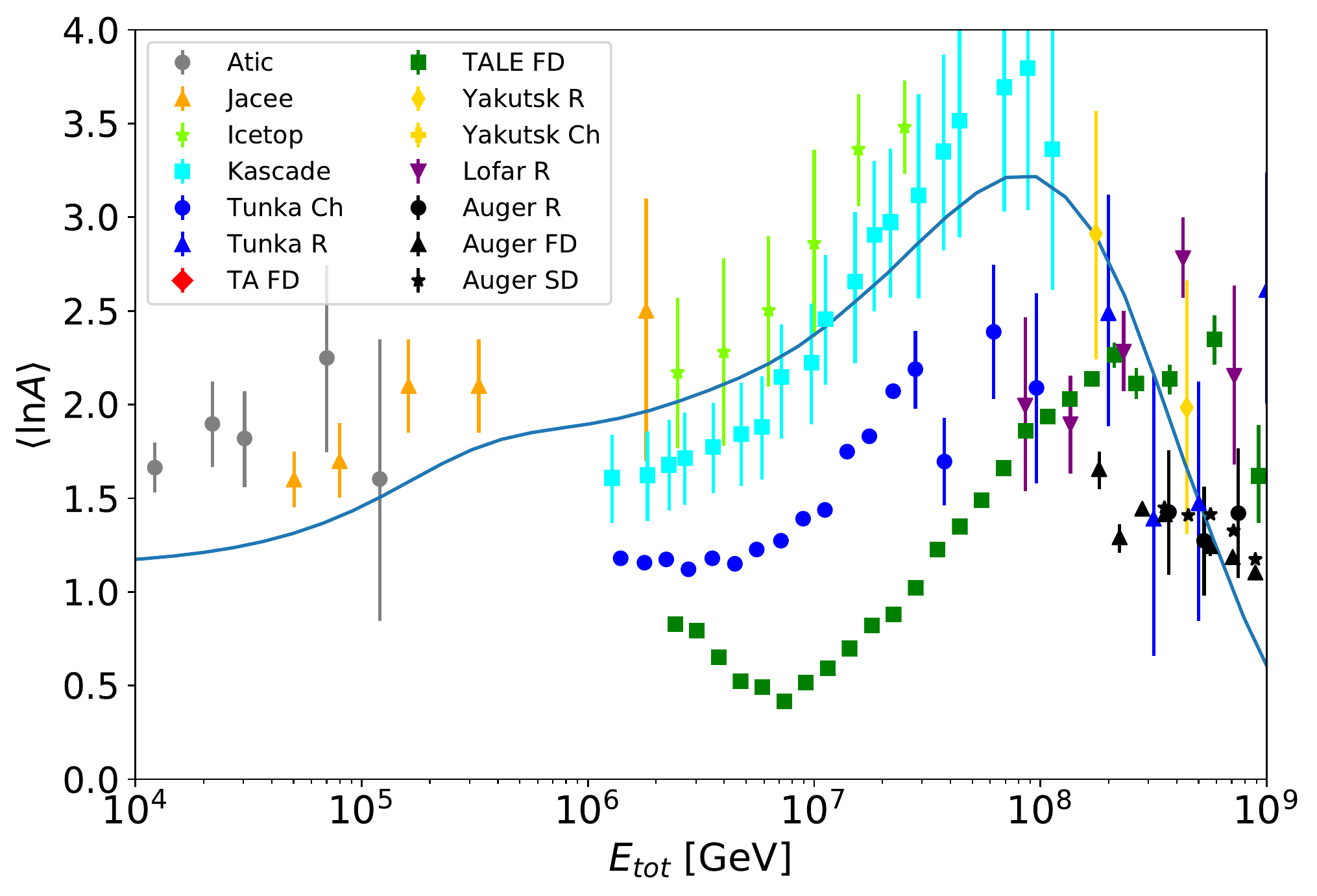}
   \caption{CR composition (mean logarithmic atomic mass) obtained from our computation, compared with recent data taken from \citet{sveshnikova2013,supanitsky2022}, and references therein.}
   \label{fig:result_composition}
 \end{figure}

Finally, we compute the mean logarithmic mass as inferred from our model and compare it with available data. The result is shown in Figure~\ref{fig:result_composition}. Once again, the variance of the data from experiment to experiment is large, due to the uncertainty in the modelling of the air shower, in particular when computing the average depth of the shower maximum of a given species. Most recent data (e.g. from the TALE or Tunka observatories) suggest that the composition around 100~PeV is lighter than previously thought (e.g. compared to older measurements by Kascade). Nevertheless, our model matches reasonably the trend of the data, and in particular the peak around 100~PeV which is due, in our scenario, to the transition between the heaviest GCRs which are the only ones which WBCSN shocks can accelerate beyond 100~PeV, and the light extragalactic component due to, e.g. the photo-disintegration of nuclei around sources \citep{unger2015,globus2015}.

\section{Conclusions}\label{sec:conclusions}
Only fast SNR shocks expanding in the collective outflow around a wind-blowing cluster might, under favourable conditions, accelerate protons beyond 10~PeV, which is required to account for the galactic component of the local CR spectrum. In this work, we have estimated the rate of these SNR shocks by analysing the massive star cluster population of the Milky Way. We found that about 15\% of the MSCs blow a collective wind. Taking into account a clustering fraction of massive stars of 80\% and a probability of 0.5 that a SN explodes close enough to the cluster edge in order to launch a fast shock in the collective wind, the rate of these events is less than 6\% that of galactic CCSN, i.e. about 1 per millennium for 3000~km/s shocks, and less than 1 per 10~kyr for 30~000~km/s shocks. We nevertheless demonstrated that this is enough to account for the spectrum of galactic CRs from TeV bands up to hundreds of PeV, where a light sub-ankle extragalactic component is expected to take over. The knee of the CR spectrum is then found to be the signature of the transition between the contribution of standard SNR shocks expanding in loose clusters (too evolved or too extended to blow a collective wind) and that of scarce SNR shocks expanding in the vicinity of wind-blowing clusters (young and compact enough to sustain a collective wind). The continuity at the knee is found to be a consequence of the second population being naturally a fraction of order $10-20\%$ the galactic CCSN population. A good agreement is obtained as well with the spectrum of the protons and data on composition.

Our model thus provides a possible explanation for GCRs, as produced in star clusters without ad-hoc hypotheses, and with a sensible set of free-parameters. Crucially, all parameter values are physically motivated and, as summarised in Table~\ref{tab:parameters}, within expected bounds. The main uncertainty of the model is the distribution of initial velocities of SN shocks. As fast SN shocks are needed to accelerate heavy elements up to hundreds of PeV, it is necessary that such a distribution displays a rather extended tail. Such a hypothesis is not excluded by available data, although the interpretation of the latter is not straightforward. Different assumptions can be set for simplicity, e.g. constant explosion energy, or constant ejected mass, which all succeed to account for the CR spectrum. Finally, it yet remains to be proven that SNR shocks expanding in cluster winds are able to accelerate protons up to several PeV, which in particular requires a high level of turbulence in the cluster, which is at the moment mostly supported by theoretical arguments.

\section*{Acknowledgements}
TV acknowledges L. Harer, S. Gabici, L.-M. Bourguinat and M. Tsirou for helpful discussions.

\section*{Data Availability}
No new data were generated or analysed in support of this research.



\bibliographystyle{mnras}
\bibliography{biblio} 






\bsp	
\label{lastpage}
\end{document}